 \documentclass[final,5p,times,twocolumn]{elsarticle}

\usepackage{amssymb}
\usepackage{amsmath}
\usepackage{hyperref}
\usepackage{mycommands}

\usepackage{array}
\usepackage{cuted}

\journal{Physics Letters B}

\begin{document}

\begin{frontmatter}

\title{Understanding parton evolution in matter from renormalization group analysis}

\author[1,2]{Weiyao Ke}
\author[1]{Ivan Vitev}
\affiliation[1]{
organization={Theoretical Division, Los Alamos National Laboratory},
addressline={Los Alamos, 87545, NM, United States}
}
\affiliation[2]{
organization={
Institute of Particle Physics and Key Laboratory of Quark and Lepton Physics (MOE), Central China Normal University
},
addressline={Wuhan, 430079, Hubei, China}
}

\begin{abstract}
We perform a renormalization group (RG) analysis of collinear hadron production in deep inelastic scattering on nuclei.  We consider the limit where the parent parton energy $E$ is large, while the medium opacity $L/\lambda_g$ remains small. We identify the fixed order and leading $\ln(E/\xi^2 L)$ enhanced medium contributions to the semi-inclusive cross sections and derive RG equations that resum multiple emissions near the endpoints of the splitting functions at first order in opacity. 
These evolution equations treat the same type of radiation enhancement in matter as the modified Dokshitzer-Gribov-Lipatov-Altarelli-Parisi approach, but differ in the way one regulates the collinear divergences. They provide a unique analytic insight into the problem of resummation and a faster and more efficient path to phenomenology. The new RG evolution framework is applied to study fragmentation in $e$A reactions.  
\end{abstract}


\end{frontmatter}

\section{Introduction}
\label{introduction}

A common characteristic of many problems in science is that microscopic fluctuations in the system manifest themselves in macroscopic effects. Such problems arise in fields ranging from high energy and nuclear physics to fluid dynamics and social networks~\cite{Newman_1999,Yankhot86,Jalilian-Marian:1997ubg,Bogner:2006pc}. They are most prevalent in divergent theories and efficiently addressed using renormalization group (RG) analysis~\cite{tHooft:1972tcz,Wilson:1973jj}. Effective theories of quantum chromodynamics (QCD) geared toward jet physics~\cite{Bauer:2000yr,Beneke:2002ph} are ideally suited to this approach and have provided new insights into renormalization and resummation.  Recently, they have given a modern perspective to the problem of parton propagation in nuclear matter~\cite{DEramo:2010wup,Ovanesyan:2011kn,Ovanesyan:2011xy,Kang:2016ofv}. These advances are essential for the interpretation of the data from reactions with nuclei at current and future colliders~\cite{AbdulKhalek:2021gbh}.    
Over the past two decades, medium-induced parton emission effects \cite{Guo:2000nz,Wang:2001ifa,Baier:2001yt,Salgado:2003gb} have been successfully implemented in jet quenching phenomenology to describe the modification of hadron and jet cross sections, and jet substructure in nuclear collisions~\cite{Qin:2015srf,Vitev:2002pf,Gyulassy:2003mc,PhysRevLett.100.072301,Wang:2009qb,Blaizot:2013hx,Kang:2014xsa,Chien:2015hda,Noronha-Hostler:2016eow,Chang:2016gjp,Zigic:2018ovr,Chen:2020tbl,Schlichting:2020lef}. Still, resummation of QCD radiation in nuclear matter remains challenging, especially lacking in analytic insight. We address this long-standing problem using RG techniques. 
If we consider semi-inclusive hadron production in deep inelastic scattering (DIS) on a nuclear target ($e$A$\rightarrow h+X$), we encounter a number of energy and length scales (defined in the target rest frame) including: 1) the hard scale $Q$, 2) the energy of the virtual photon (jet) $\nu(E)$, 3) the path length $L$, 4) the mean free path $\lambda_{\rm g}$ and 5) the inverse interaction range $\xi$. $\xi$ is also the effective parton mass that arises in the soft background gluon field of the nucleon~\cite{Qiu:2003vd,Qiu:2004da}.
Therefore,  observables in $e$A are functions of many dimensionless  parameters
\begin{align}
{\rm Obs} \equiv {\rm Obs}\left(\frac{Q}{Q_0}, \frac{E}{\xi^2 L}, \frac{L}{\lambda_g}, \lambda_g \xi, \frac{\xi}{Q_0}, \cdots\right).
\label{eq:obs}
\end{align}
A simplified description with controlled accuracy is often possible when one (or more) of these dimensionless ratios become asymptotically large \cite{barenblatt_1996}. 
For example, the limit $\lambda_g\xi \gg 1$ 
allows the use of time-ordered perturbation theory to derive quark and gluon splitting functions in matter~\cite{Ovanesyan:2011xy,Sievert:2019cwq}. A partonic transport picture emerges when  $L/\lambda_g\gg 1$ in a thick and dense medium~\cite{Schenke:2009gb,Blaizot:2013vha,He:2015pra}.

In this letter, we compute hadron production in the regime where $Q/Q_0$, $E/\xi^2 L, \lambda_g\xi$ become asymptotically large, while $L/\lambda_g, \xi/Q_0$ stay at order unity/few.
This limit is particularly relevant for high-energy hadron production in thin, dilute, or fast-expanding media. 
To resum the large $\ln (Q/Q_0)$ and $\ln (E/\xi^2L)$ enhancements from the vacuum and medium-induced radiative corrections,
we  introduce two final-state renormalization scales $\mu_1$ and $\mu_2$ in the single-inclusive hadron cross section~\cite{deFlorian:1997zj}
\begin{align}   
 \frac{d\sigma_{eA\rightarrow h}}{dx_B dQ^2 dz_h} &=  \frac{2\pi\alpha_{e}^2 }{Q^4}  \sum_{i,j} e_q^2 \Big\{\left\{f_{j/A} \otimes \mathcal{C}_{ij}\right\}_{x_B}  \otimes d_{h/i} \Big\}_{z_h}\, , \label{eq:SIDIS}\\
\mathcal{C}_{ij}(x,z) &= \left[(1+(1-y)^2)\mathcal{C}_{ij}^{1}+2(1-y)\mathcal{C}_{ij}^{L}\right] \, , \\
 \left\{ h\otimes g\right\}_x &\equiv \int_{x}^1  h\left(\frac{x}{x'}\right) g(x')\frac{dx'}{x'}\,  \nonumber . 
\end{align}
Here, $y=\nu/E_e$ with $E_e$ and $\nu=Q^2/(2 x_B M_p )$ are the energies of the incident electron and the virtual photon, respectively. 
$f_{j/A}(x, Q^2)$, $ d_{h/i}(z, \mu_1^2, \mu_2^2)$ and $\mathcal{C}_{ij}^{1,L}(x, z, Q^2, \mu_1^2, \mu_2^2)$ are the parton distribution functions (PDFs), fragmentation functions (FFs), and the hard coefficient functions with fractional electric charge $e_q$.
The PDFs are evaluated at scale $Q^2$, and the dependence on $Q^2$ will not be written explicitly hereafter.
Since the cross-section $\sim f_{j/A}\otimes \mathcal{C}_{ij}^{1,L} \otimes d_{h/i}$ must not depend on $\mu_1$ and $\mu_2$, it is sufficient to study the scale evolution of $F_{ij}^{1,L}(z)\equiv f_{j/A}\otimes z\mathcal{C}_{ij}^{1,L}$, while $d_{h/i}$ will obey the same evolution but with the opposite sign.
This quantity can be interpreted as the invariant distribution of parton $i$ produced by the hard scattered parton $j$,
 allowing us to track the evolving energy $E=z\nu$. 
At tree level $F_{ij}^1= z\delta_{ij}f_{j/A}(x_B)\delta(1-z)$, $F_{ij}^L = 0$, and we work up to next-to-leading order (NLO) with $\mathcal{C}_{ij}^{1,L}$ taken from~\cite{deFlorian:1997zj,Furmanski:134618}. 

As will become clear in a moment, the $Q/Q_0$ and $E/\xi^2L$ enhancements have distinct physics origins.
Therefore, in addition to the vacuum renormalization that leads to  DGLAP evolution, the ``medium bare'' $F_{ij}$ needs to be further renormalized by a medium coefficient $M_{ij}$ that only depends on $\mu_2$,
\begin{align} 
F_{ij}(z,\mu_1^2,\mu_2^2)&\rightarrow M_{ik}\left(z/y, \mu_2^2\right)\otimes F_{kj}(y,\mu_1^2, \mu_2^2)   +\mathcal{F}_{ij}(z) ,
\label{eq:factorized-assumption}\\
M_{ik} &= M_{ik}^{(0)} + M_{ik}^{(1)}+\cdots ,
\end{align}
where $M_{ik}^{(0)}=y\delta_{ik} \delta(1-y)$ and the first non-trivial counter term is $M_{ik}^{(1)}$.
$\mathcal{F}_{ij}(z)$ are medium corrections that are subleading in $\ln (E/\xi^2 L)$, i.e. other fixed order (FO) medium contributions.

\section{Renormalization group analysis}
We consider corrections to $F_{ij}(z)$ from both vacuum and medium-induced collinear splittings, which have the form
\begin{align}
\Delta F_{ij} = \frac{\alpha_s^{(0)}}{2\pi^2} \int^{Q^2} \frac{  d^{2-2\epsilon} \bfk}{\bfk^2 } x[P_{ik}]_+\otimes F_{kj}  + \Delta F_{ij}^{\rm m}(z,\mu_2) \, .
\label{eq:naive-nlo-vac+med}
\end{align}
We work in $d=4-2\epsilon$ dimensions. Here, $\alpha_s^{(0)}$ is the bare coupling, which is related to the leading-order running coupling constant by $\alpha_s^{(0)} = (4\pi)^{-\epsilon}e^{\gamma_E\epsilon} \mu^{2\epsilon} \alpha_s(\mu^2)$. $P_{ik}$ are the vacuum Altarelli-Parisi splitting functions~\cite{Altarelli:1977zs,Lipatov:1974qm}, with the ``+" prescription in Eq.~(\ref{eq:naive-nlo-vac+med}) only applied to diagonal contributions.
In the first term, the renormalization scale $\mu=\mu_1$ acts as a collinear cutoff of vacuum emissions, in which case RG analysis leads to DGLAP evolution of $F_{ij}(z)$ that resums $\ln(Q/Q_0)$.

\begin{table}[b!]
\centering
\begin{tabular}{|>{\centering\arraybackslash}p{1cm}|>{\arraybackslash}p{1.8cm}>{\arraybackslash}p{1.8cm}>{\arraybackslash}p{2.5cm}|}
\hline
$j\rightarrow i$ & $C_1^{ij},~(\Delta_1^{ij})^2$ & $C_2^{ij},~(\Delta_2^{ij})^2$ & $C_3^{ij},~(\Delta_3^{ij})^2$\\
\hline
$q\rightarrow q$ & $C_A,~x^2$ & $C_A,~1$ & $2C_F-C_A,~(1-x)^2$ \\
$q\rightarrow g$ & $C_A,~1$ & $C_A,~(1-x)^2$ & $2C_F-C_A,~x^2$ \\
$g\rightarrow q$ & $C_A,~(1-x)^2$ & $C_A,~x^2$ & $2C_F-C_A,~1$ \\
$g\rightarrow g$ & $C_A,~1$ & $C_A,~x^2$ & $C_A,~(1-x)^2$ \\
\hline
\end{tabular}
\caption{Color ($C_n^{ij}$) and kinematic factors ($\Delta_n^{ij}$) in Eq.~(\ref{eq:collinear-pm-s}).}
\label{table:T1}
\end{table}

Next, we extract the leading logarithmic and fixed-order contributions from the medium-induced correction in the second term of
Eq.~(\ref{eq:naive-nlo-vac+med}).
For thin and uniform nuclear matter of length $L$ we start with the splitting functions $P_{ij}^{(1)}(x)$~\cite{Ovanesyan:2011xy,Ovanesyan:2015dop,Kang:2016ofv, Sievert:2019cwq}  obtained in the opacity expansion approach using soft-collinear effective theory with Glauber gluons  (SCET$_{\rm G}$)~\cite{DEramo:2010wup,Ovanesyan:2011kn}. 
The full expressions still contain multiple scales, including $E/L, \xi^2$, and $Q^2$ as integration boundaries in the phase space for parton branching. Thus, they are not yet suitable for RG analysis. To rectify this, for a medium of uniform density the line integral along the path of parton propagation can be performed. Neglecting power-suppressed terms under dimensionally regularized integration over the transverse momenta of the splitting $\bfk$ and the Glauber gluon exchange $\bfq$, we change integration variables to arrive at an intermediate expression  
\begin{align}
&P_{ij}^{(1)}(x,E, \mu_2^2) =  \frac{\alpha_s^{(0)} P_{ij}(x)}{2\pi^2} L \int \frac{d^{2-2\epsilon}\bfk}{(2\pi)^{-2\epsilon}}\frac{\Phi\left(\frac{\bfk^2 L}{2x(1-x)E}\right)}{\bfk^2}  \nonumber \\  
&\quad\quad\sum_n \int \frac{d^{2-2\epsilon}\bfq}{(2\pi)^{-2\epsilon}} 
\frac{\alpha_s^{(0)}}{\pi}\rho_G\frac{  C_{n}^{ij}\Delta_n^{ij}(x)}{ (\bfq^2+\xi^2)^2} \frac{\bfq\cdot[\bfk+\Delta_n^{ij}(x) \bfq]}{[\bfk+\Delta_n^{ij}(x) \bfq]^2}  \nonumber \\
&= \frac{\alpha_s^2(\mu_2^2)\rho_G L}{8E/L}\frac{P_{ij}(x)}{[x(1-x)]^{1+2\epsilon}}
\left[\frac{e^{\gamma_E}\mu_2^2}{2E/L}\right]^{2\epsilon}  \int_0^{w}  du \frac{4}{\pi} \frac{\Phi(u)}{u^{1+\epsilon}}  \nonumber \\  
&\frac{\epsilon\Gamma(\epsilon)}{ \Gamma(1-\epsilon)}\sum_{n}C_{n}^{ij}[\Delta_n^{ij}]^{2-2\epsilon} \int_0^1 \frac{ds}{(1-s)^{\epsilon}} \frac{\frac{[\Delta_n^{ij}]^2}{2x(1-x)}\frac{1-\epsilon}{2}\delta-\epsilon su}{\left[su+\frac{[\Delta_n^{ij}]^2}{2x(1-x)}\delta\right]^{2+\epsilon}}
\label{eq:collinear-pm} \, .
\end{align}
Here, $\Phi(u) = 1-\sin(u)/u$ and the color and kinematic factors $C_n^{ij}$ and $\Delta_n^{ij}(x)$ are collectively defined in Table~\ref{table:T1}. 
The coupling constant, color factors, and the density of scattering centers in the target (T) are absorbed in an effective Glauber gluon density  
\begin{align}
\rho_G &= \sum_T  \rho_T 4\pi\alpha_s^{\rm med} C_T/d_A \, . \;
\end{align}
The reasoning behind this redefinition is that radiative correction to the medium parton and coupling cannot change the collinear parton at leading power and, therefore, will lead to scaleless integrals that vanish in an effective theory. This means that for collinear observables $\rho_G$ does not receive NLO corrections. It will be used as an effective non-perturbative parameter. 
The relation between our cold nuclear matter parameters $\rho_G$ and $\xi^2$ with the conventionally used jet transport parameter is
\begin{align}
\hat{q}_R &= \rho_G \int \frac{\alpha_s C_R }{\pi}\frac{\bfq^2}{(\bfq^2+\xi^2)^2}d^2\bfq.
\end{align}

In the second line of Eq.~(\ref{eq:collinear-pm}) we have performed the $\bfq$ integration, with $s$ being a Feynman parameter. The remaining $\bfk$ integration is expressed in the dimensionless variable $u=\bfk^2 L/2x(1-x)E$, and $w=Q^2L/2\nu$ is used as the upper bound of the $u$ integration. 
To finally reduce in-medium splittings to a single-scale problem in the large separation of scales limit $E/L\gg \xi^2$, we expand in the small parameter $\delta = \xi^2 L/E \ll 1$, consistent with the collinear power counting. Then, at the leading power in $\delta$, one can complete the $u$ integration exactly to arrive at a single-scale generic form
\begin{align}
 P_{ij}^{(1)}(x,E, \mu_2^2) &=    \frac{\alpha_s^2(\mu_2^2) B(w) \rho_G L}{8E/L} \frac{ P_{ij}(x)}{[x(1-x)]^{1+2\epsilon}} \left[\frac{\mu_2^2  L}{\chi(w) E}\right]^{2\epsilon} \nonumber\\
 &\sum_n C_n^{ij}[\Delta_n^{ij}(x)]^{2-2\epsilon}\left[1+\mathcal{O}\left(\epsilon^2\right)\right] \left[1+\mathcal{O}(\delta)\right]  \, .   
\label{eq:collinear-pm-s}
\end{align}
The medium-induced splitting functions in Eq.~(\ref{eq:collinear-pm-s}) only contain a single scale $E/L$. $\mu_2^2$ is an evolution scale that is limited by $\xi^2$ from below. Later, we will see that this behavior leads to a collinear logarithmic factor $\ln(L/\chi E \xi^2)$.  We also note that $P_{ij}^{(1)}$ acquires additional $[x(1-x)]^{-1-2\epsilon}$ singularities as compared to $P_{ij}$. 
We do not assume any ordering between $Q^2$ and $E/L$ as there is no additional logarithmic enhancement associated with the limit $Q^2 \gg E/L$. The $Q^2$ dependence is encoded in two well-behaved coefficient functions $B(w)$ and $\chi(w)$
\begin{align}
B(w) &= \frac{4}{\pi}\int_0^{w}  \frac{\Phi(x)}{x^2} dx \, , \\
\chi(w) &= 2\exp\left\{\frac{1}{B(w)}\frac{4}{\pi}\int_0^{w}\frac{\Phi(x) }{x^2} \ln(x) dx \right\} \, ,
\end{align}
via $w=Q^2L/2\nu$. For the SIDIS process at moderate Bjorken $x_B$, $w \equiv x_B M_p L \approx 6.0 x_B A^{1/3}$ is of order few. For fragmentation at midrapidity in hadronic collisions, $w \sim EL \rightarrow \infty$, and $B(\infty)=1$, $\chi(\infty)=2e^{3/2-\gamma_E}\approx 5.03$.
This analytic approach allows us to integrate out the space-time information in a finite medium, properly accounting for the non-local coherent nature of in-medium parton branching, which cannot be achieved in classical time-ordered Monte Carlos. 

With  Eq.~(\ref{eq:collinear-pm-s}) at hand, taking the medium-induced $q\rightarrow q$ contribution in Eq.~(\ref{eq:naive-nlo-vac+med}) as an example,
\begin{align}
& \Delta F_{qj}^{\rm m}(z,\mu_2^2) = \left(M_{qq}^{(0)} + M_{qq}^{(1)}\right)\otimes {\left[P^{(1)}_{qq}\right]_+} \otimes F_{qj}  \nonumber \\
&\approx \int_0^1 dx \left[P^{(1)}_{qq}\left(x, \frac{z}{x}\nu, \mu_2^2\right) F_{qj}\left(\frac{z}{x}\right) -P^{(1)}_{qq}\left(x, z\nu, \mu_2^2\right)F_{qj}\left(z\right)\right] \nonumber\\
&+ \int_0^1 \frac{dx}{x} M_{qq}^{(1)}\left(\frac{z}{x}, \mu_2\right) F_{qj}(x)  \, ,
\label{eq:naive-nlo}
\end{align}
where we have used the tree level $M_{qq}^{(0)}$, while the NLO counter term $M_{qq}^{(1)}$ will be determined after we identify the $1/\epsilon$ poles of the first term.
To achieve this, we note that $P_{ij}^{(1)}(x)$  
in Eq.~(\ref{eq:collinear-pm-s}) contain additional $[x(1-x)]^{-1-2\epsilon}$ divergences as compared to the vacuum splitting functions, which do not cancel among real and virtual corrections in Eq.~(\ref{eq:naive-nlo}).
Taking the $q\rightarrow q$ channel as an example, we can isolate these $x=0,1$ divergences, including the multiplicative $(1-x)^{-1}$ contribution from the vacuum $P_{qq}(x)$, and separate the resulting $1/\epsilon$ poles in convolution with any well-behaved function $G(x)$,
using the following decomposition 
\begin{align}
\int_0^1 \frac{G(x)}{x^{1+2\epsilon}(1-x)^{2+2\epsilon}} dx &= \int_0^1 \frac{\left\{G(x)\right\}_{qq}}{x(1-x)^2 }dx\nonumber\\
-\frac{G(0)}{2\epsilon} - G&(1)\left(\frac{1}{2\epsilon}+2\right) +\frac{G'(1)}{2\epsilon} + \mathcal{O}(\epsilon) \, .
\label{eq:sub1}
\end{align}
The result has been expanded near $\epsilon=0$, and the curly bracket denotes the subtracted function
$\left\{G(x)\right\}_{qq} = G(x)-(1-x)^2 G(0)
- x(2-x)G(1) - x(x-1)G'(1). $
It is straightforward to check that the integral on the right-hand side is free from divergences at $x=0,1$. The second line of Eq.~(\ref{eq:sub1}) contains all singularities. Due to the double pole at $x=1$ one needs to subtract both $G(x)$ and the derivative $G'(x)$ at $x=1$, and such derivative subtractions (higher-order ``plus'' prescriptions) are also used to treat subleading power corrections in SCET~\cite{Ebert:2018gsn}. 
Similarly, we can separate the $1/\epsilon$ poles in convolution with $P^{(1)}_{gg}(x)$ by
\begin{align}
\int_0^1 \frac{G(x)}{x^{2+2\epsilon}(1-x)^{2+2\epsilon}} &dx = \int_0^1 \frac{\left\{G(x)\right\}_{gg}}{x^2(1-x)^2 }dx  \nonumber\\  -\left[G(0)+G(1)\right]&\left[\frac{1}{\epsilon}+2\right]-\frac{G'(0)}{2\epsilon} + \frac{G'(1)}{2\epsilon}+ \mathcal{O}(\epsilon)  \, ,  \label{eq:sub2} 
\end{align}
with $\left\{G(x)\right\}_{gg} = G(x)-(1-x)^2\left[(1+2x)G(0)+xG'(0)\right] - x^2\left[(3-2x)G(1)+(x-1)G'(1)\right]$. For $P^{(1)}_{qg}(x)$, this relation is
\begin{align}
\int_0^1 \frac{G(x)}{x^{1+2\epsilon}(1-x)^{1+2\epsilon}} dx &= \int_0^1 \frac{\left\{G(x)\right\}_{qg}}{x(1-x) }dx  \nonumber\\  &-\frac{G(0)}{2\epsilon} -\frac{G(1)}{2\epsilon}  + \mathcal{O}(\epsilon)   \, , \label{eq:sub3}
\end{align}
with $\left\{G(x)\right\}_{qg} = G(x)- x G(1) - (1-x)G(0)$. 
Finally, the corresponding formula for convolution with $P_{gq}^{(1)}$ can be obtained by applying $x\rightarrow 1-x$ transformation to Eq.~(\ref{eq:sub1}).
Such relations allow us to separate $1/\epsilon$ poles from the convolution in Eq.~(\ref{eq:naive-nlo}).

Following  Eq.~(\ref{eq:sub1}), the $q\rightarrow q$ contribution to the flavor non-singlet sector $\Delta F_{\rm NS} = \Delta F_{qj}^m - \Delta F_{\bar{q}j}^m$ can be decomposed into $1/\epsilon$ poles and logarithmic enhanced terms, counter terms, and fixed-order contributions. Using the relation $E=z\nu$, we find
\begin{align}
 \Delta F_{\rm NS} &=   \frac{\alpha_s^2(\mu_2^2) B(w)\rho_G L}{8\nu/L}  \left(\frac{1}{2\epsilon}+\mathcal{L}\right)\nonumber\\
 & \times 2C_F\left(\frac{2C_A+C_F}{z} - 2 C_A \frac{d}{dz}\right)F_{\rm NS}(z) \nonumber\\
 & + M_{qq}^{(1)} \otimes F_{\rm NS} + \Delta \mathcal{F}_{NS} \, . 
\label{eq:NLO-decomposed}
\end{align}
The logarithmic factor $\mathcal{L} = \ln \left(\mu_2^2L/\chi(w) z\nu\right)$ and $\Delta\mathcal{F}_{\rm NS}$ stands for non-logarithmic-enhanced fixed-order contributions.
The leading $L/\nu = 2 m_N x_B L/Q^2$ dependence in Eq.~(\ref{eq:NLO-decomposed}) shows that these are higher twist terms, but enhanced by the nuclear size.
The medium contribution has a natural scale $\mu_2^2= \chi(w) z \nu/L$. 
Divergences associated with the $P_{qq}(x)$ factor in $P_{qq}^{(1)}(x)$ have canceled among the real and virtual terms. 
The remaining poles come from extra $x\rightarrow 0,1$ divergences of the medium-induced emission spectra.
Similarly, using Eqs. (\ref{eq:sub1}), (\ref{eq:sub2}) and (\ref{eq:sub3}) medium corrections to the flavor-singlet quark $\Delta F_{f} = \Delta F_{qj}^m + \Delta F_{\bar{q}j}^m$  and gluon $\Delta F_{g} = \Delta F_{gj}^m$ sectors are
\begin{align}
\Delta F_g &= \frac{\alpha_s^2(\mu_2^2) B(w)\rho_G L}{8\nu/L}\left(\frac{1}{2\epsilon}+\mathcal{L}\right)\nonumber\\
&\times \left[- 4C_A^2 \frac{dF_g(z)}{dz} + C_F   \frac{2N_fF_g(z)- 2C_F \sum_f F_f(z)}{z}\right] \nonumber\\
 & +  M_{gg}^{(1)} \otimes F_{g} +  M_{gq}^{(1)} \otimes \sum_f F_{f} + \Delta \mathcal{F}_{g} \,  , 
\\
\Delta F_f &= \frac{\alpha_s^2(\mu_2^2) B(w)\rho_G L }{8\nu/L}\left(\frac{1}{2\epsilon}+\mathcal{L}\right)\nonumber\\
&\times 2C_F\left[\left(\frac{2C_A+C_F}{z}- 2 C_A \frac{d}{dz}\right)F_{f}(z) -T_R  \frac{F_g(z)}{z}\right] \nonumber\\
& + M_{qq}^{(1)} \otimes  F_{f} + 2M_{qg}^{(1)} \otimes F_{g} + \Delta \mathcal{F}_{f} \, . 
\end{align}
We can now define the NLO medium counter terms $M_{ij}^{(1)}$ such that they cancel the $1/\epsilon$ poles in the calculation to this order~\footnote{We have checked the same $1/\epsilon$ divergences with an opposite sign arise from the soft-collinear modes. Thus, the  cancellation of the remaining infrared divergences and the counter terms are associated with this sector of the theory.}.

Finally, we list fixed-order (FO) corrections that are free of divergences and large logarithms. 
For the flavor non-singlet sector,
\begin{align}
\label{eq:foNS}
& \Delta\mathcal{F}_{\rm NS} = \frac{\alpha_s^2(\chi z\nu/L) B(w) \rho_G L}{8\nu/L}\left\{\frac{(4C_A-C_F)C_FF_{\rm NS}(z)}{z} \right.\\
&\left.+\int_0^1 \frac{\left\{ \sum_n  C_n^{qq}[\Delta_n^{qq}(x)]^2C_F (1+x^2)\left[ \frac{x}{z}F_{\rm NS}\left(\frac{z}{x}\right)-\frac{F_{\rm NS}(z)}{z}\right]\right\}_{qq}}{x(1-x)^2}dx \right\} . \nonumber
\end{align}
For the flavor singlet sector, they are
\begin{align}
\label{eq:foG}
& \Delta \mathcal{F}_g = \frac{\alpha_s^2(\chi z\nu/L) B(w) \rho_G L}{8\nu/L}\left\{14 C_A^2 \frac{F_g(z)}{z} \right. \\
& \left. 
-3C_F^2  \sum_f \frac{F_f(z)}{z}+ N_f T_R \frac{4}{3}(C_A+3C_F) \frac{F_{g}(z)}{z}\right.\nonumber\\
& + \sum_f \int_0^1 dx \frac{\left\{\sum_n C_n^{gq}[\Delta_n^{gq}]^2 C_F[1+(1-x)^2]\frac{x}{z}F_f(\frac{z}{x})\right\}_{gq}}{x^2(1-x)} \nonumber\\
&\left.+ \int_0^1 dx \frac{\left\{\sum_n C_n^{gg}[\Delta_n^{gg}]^2 C_A2\left(1-x+x^2\right)^2\left[\frac{x}{z}F_g(\frac{z}{x})-x \frac{F_g(z)}{z}\right]\right\}_{gg}}{x^2(1-x)^2} \right\} \nonumber ,\\
\label{eq:foSS}
& \Delta \mathcal{F}_f = \frac{\alpha_s^2(\chi z\nu/L) B(w) \rho_G L}{8\nu/L}\left\{ (4C_A-C_F)C_F \frac{F_{f}(z)}{z} \right.\\
&+\int_0^1 \frac{\left\{\sum_n  C_n^{qq}[\Delta_n^{qq}(x)]^2 C_F(1+x^2)\left( \frac{x}{z}F_{f}\left(\frac{z}{x}\right)-\frac{F_{f}(z)}{z}\right)\right\}_{qq}}{x(1-x)^2}dx \nonumber\\
& \left.+ \int_0^1 dx \frac{\left\{\sum_i C_n^{qg}[\Delta_n^{qg}]^2T_R[x^2+(1-x)^2]\frac{x}{z}F_g(\frac{z}{x})\right\}_{qg}}{x(1-x)} \right\}\nonumber .
\end{align}

\subsection{The in-medium RG evolution}

With the $1/\epsilon$ pole cancelled by the counter term, we take a derivative with respect to $\ln \mu_2^2$ on both sides of Eq.~(\ref{eq:NLO-decomposed}) and keep leading terms in $\alpha_s$ to obtain an RG equation for $F_{\rm NS} = F_{qj}-F_{\bar{q}j} $ quark spectra
\begin{align}
\frac{\partial F_{\rm NS}(\tau, z)}{\partial \tau} &= 2C_F\left( 2C_A   \frac{\partial}{\partial z} - \frac{2C_A+C_F}{z} \right) F_{\rm NS} \, , \; \; 
\label{eq:main-eq-NS}\\
\tau(\mu_2^2) &= \frac{ B(w)\rho_G L}{8\nu/L} \frac{4\pi}{\beta_0} \left[\alpha_s(\mu_2^2) - \alpha_s\left(\frac{\chi(w) z\nu}{L}\right)\right],
\end{align}
where we define a new variable $\tau(\mu_2^2)$ for the scale evolution to take into account the running coupling effect, with $\beta_0 = (11  -2N_f/3)$, and absorb medium parameters and numerical factors.
Similarly, the equations for the flavor-singlet sector are 
\begin{align}
\label{eq:main-eq-S1}
\frac{\partial F_f}{\partial\tau} &= 2C_F\left( 2C_A \frac{\partial}{\partial z} - \frac{2C_A+C_F}{z} \right) F_f + C_F \frac{F_g}{z}\,  ,  \;  \;  \;  \\
\frac{\partial F_g}{\partial \tau} &= \left(4C_A^2 \frac{\partial }{\partial z} - \frac{2N_f C_F}{z}\right)F_g  + 2C_F^2  \sum_{f}\frac{F_f}{z} \,  ,  \;  \;  \; 
\label{eq:main-eq-S2}
\end{align}
where $F_g\equiv F_{gj}$ is the gluon spectrum and $F_f=F_{qj}+F_{\bar{q}j}$ for $f=u,d,s$ are flavor-singlet quark spectra.
Eqs. (\ref{eq:main-eq-NS}), (\ref{eq:main-eq-S1}) and (\ref{eq:main-eq-S2})  are the main results of this letter.
Starting with an initial condition at $\mu_2^2 = \chi z \nu/L$ and evolving down to $\mu_2^2 = \xi^2$ where the in-medium interactions are cut off, the  
non-singlet Eq.~(\ref{eq:main-eq-NS}) has a very elegant traveling wave solution
\begin{align}
F_{\rm NS}(\tau, z) = \frac{F_{\rm NS}\left(0, z+ 4C_F C_A\, \tau\right)}{(1+ 4C_F C_A \, \tau/z)^{1+C_F/(2C_A)}} \, .
\label{eq:pocket}
\end{align}
The main effect of the evolution is to shift the initial distribution of partons by $\Delta z = -4C_F C_A\tau$, thus the in-medium parton energy loss can be directly obtained from our RG analysis. 
Note that in the region $z>1-4C_FC_A\tau$ the shift in the argument leads to $F(\tau, z)=0$, which seems unphysical. This is because collinear power counting breaks down when $z\approx 1$ and suggests that the traveling wave solution is not applicable in the $z=1$ threshold region.
The denominator arises from the collinear quark-to-gluon conversion, which further reduces the magnitude of the non-singlet distribution.
If off-diagonal quark-gluon coupling terms are neglected in flavor-singlet Eqs. (\ref{eq:main-eq-S1}), (\ref{eq:main-eq-S2}), similar traveling wave solutions can be obtained. For applications in this letter, we will solve them numerically in full generality.

\subsection{Connection to modified DGLAP equations}

A widely used phenomenological approach for parton evolution in matter is based on the modified DGLAP (mDGLAP) framework~\cite{Wang:2009qb,Chang:2014fba,Chien:2015vja}. Again, for the non-singlet sector
\begin{align}
\frac{\partial F_{\rm NS}(z)}{\partial \ln \mu^2} &= \int_0^1 \bfk^2 \left[\frac{ dP_{qq}(x,\bfk^2)}{d \bfk^2} + \frac {dP_{qq}^{(1)} (x,\bfk^2)}{d\bfk^2}  \right] \nonumber \\
&\times\left[F_{\rm NS}\left(z/x\right)-F_{\rm NS}(z)\right] dx \, , \; \; 
\label{eq:integal-mDGLAP}
\end{align}
with $\mu^2=\bfk^2/x(1-x)$  the virtuality of the parton, and $dP_{qq}/d\bfk^2  + dP_{qq}^{(1)} /d\bfk^2$ the differential splitting function including both vacuum and medium-induced contributions. Building upon the new analytic insights,   we can now show that the mDGLAP approach resums the same medium enhancement as the in-medium RG equations to leading logarithmic accuracy, albeit without proper separation of fixed order terms.  Unlike the RG analysis that uses dimensional regularization,  mDGLAP evaluates the full splitting functions numerically and regulates the endpoint divergences of $P_{ij}^{(1)}$ 
such that $x,1-x \geq\xi^2/\mu^2$~\cite{Ke:2022gkq}. If we focus on the medium-induced contributions near $x = 1$ and use a fixed coupling for simplicity,  Eq.~(\ref{eq:integal-mDGLAP}) becomes
\begin{align}
\frac{\partial F_{\rm NS}}{\partial \ln \mu^2} &= 4C_F C_A A_0 \int_0^{1-\frac{\xi^2}{\mu^2}} \frac{4\Phi(u)}{\pi u}  
\frac{\frac{x}{z}F_{\rm NS}(\frac{z}{x})-\frac{F_{\rm NS}(z)}{z}}{(1-x)^2} dx  \nonumber \\
& \approx \frac{4\Phi(u)}{\pi u} \, 4C_F C_A A_0 \left[\frac{\partial F_{\rm NS}}{\partial z}-\frac{F_{\rm NS}}{z}\right] \ln \frac{\mu^2}{\xi^2} \nonumber \\
 &\approx \delta \left(\mu^2-\frac{2\pi E}{L} \right) \,  4C_F C_A A_0 \left[\frac{\partial F_{\rm NS}}{\partial z}-\frac{F_{\rm NS}}{z}\right] \ln \frac{\mu^2}{\xi^2} \, , 
\label{eq:expand-mDGLAP}
\end{align}
with $u=\mu^2L/2E$ and $A_0=\alpha_{s, \rm fix}^2 \rho_G L^2 B(w)/(8\nu)$.
In the second line, 
we have performed a Taylor expansion of $(x/z)F(z/x)$ near $x=1$ and omitted subleading terms in $\ln (\mu^2/\xi^2)$.
Its partial connection to first-principles RG analysis results is most easily illustrated by considering a specific case with $\xi^2 \ll E/L \ll  Q^2$.
Because $\Phi(u)/u$ peaks at $u=\pi$ and $\int_0^\infty  \frac{4\Phi(u)}{\pi u}d\ln u=1$, one can make an impulse approximation shown in the third line of Eq.~(\ref{eq:expand-mDGLAP}). 
Then, to leading-log accuracy, one can perform vacuum DGLAP evolution above and below the scale $\mu^2=2\pi E/L$.
Under this approximation, the solution just below this scale ($F_{NS}^-$) and the solution above ($F_{NS}^+$) are related by
\begin{align}
F_{NS}^+(z) = \frac{F_{NS}^-(z+4C_FC_A \, \tau_{\rm fix})}{1+4C_FC_A \,  \tau_{\rm fix}/z } \, ,
\end{align}
with $\tau_{\rm fix} = A_0 \ln \frac{2\pi E}{\xi^2 L}$. This is the same traveling wave solution as in Eq.~(\ref{eq:pocket}), but with fixed coupling and neglecting contributions from the $x=0$ endpoint.
We are thus able to show explicitly for the first time what type of medium-induced large logarithms the mDGLAP approach with the choice $\mu^2 = \bfk^2/x(1-x)$ resums. Of course, for large scale separations, it becomes computationally intensive to evaluate the RHS of Eq.~(\ref{eq:integal-mDGLAP}) with an explicit cut-off. The RG approach that we formulated is not only more illuminating but also easier to implement. 
In fact, scale separation is key to obtaining semi-analytic and closed-form solutions to RG equations.

\begin{figure}[ht!]
    \centering
    \includegraphics[width=\columnwidth]{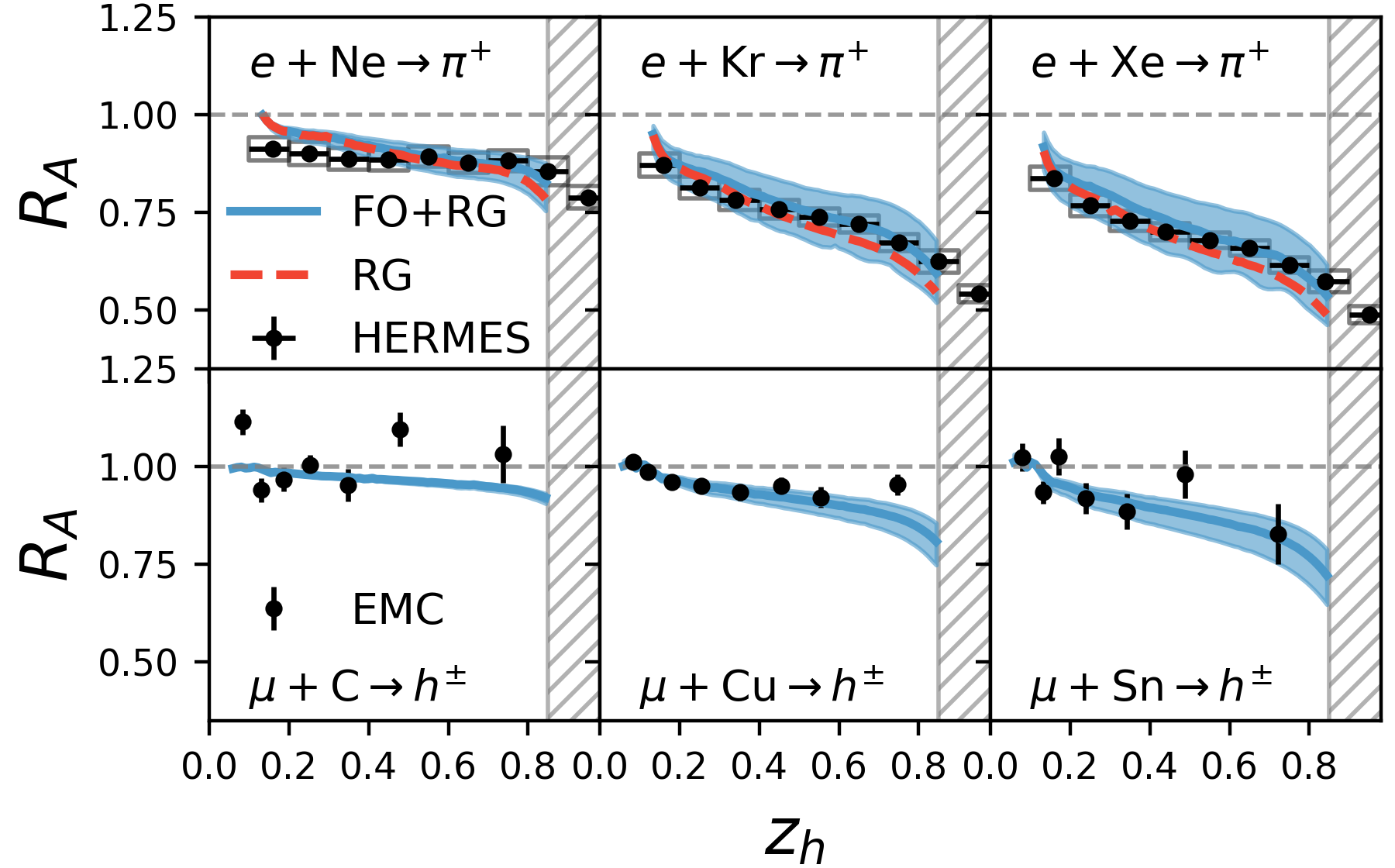}
\vspace*{-2em}
    \caption{Top panel: medium modifications to the $\pi^+$ fragmentation distributions are compared to HERMES data~\cite{HERMES:2007plz}, characterized by the average $\nu=12$~GeV, $Q^2=2.25$~GeV$^2$. Bottom panel: RG+FO calculation compared to EMC data~\cite{EuropeanMuon:1991jmx} with $\nu=62$~GeV, $Q^2=11$~GeV$^2$, and the same CNM parameters that describe the HERMES data.}
    \label{fig:RA_HERMES_EMC}
\end{figure}

\section{Phenomenology}
 We demonstrate the new method by studying nuclear effects on single-hadron fragmentation in SIDIS given by Eq.~(\ref{eq:SIDIS}). 
We implement the fully-coupled RG evolution Eqs.~(\ref{eq:main-eq-NS}), (\ref{eq:main-eq-S1}), (\ref{eq:main-eq-S2}), and the fixed order terms from Eqs.~(\ref{eq:foG}), (\ref{eq:foNS}), (\ref{eq:foSS}).
The nuclear modification is defined as the ratio of inclusive-normalized SIDIS cross sections between electron-nucleus ($e$A) and electron-deuterium ($ed$ for HERMES and EMC) or electron-proton ($ep$ for EIC) collisions
\begin{align}
R_A(x_B, Q^2, z_h) = \frac{\left.\frac{d\sigma_{eA\rightarrow h}}{dx_B dQ^2 dz_h}\right/\frac{d\sigma_{eA}}{dx_B dQ^2}}{ \left.\frac{d\sigma_{ed,ep\rightarrow h}}{dx_B dQ^2 dz_h}\right/ \frac{d\sigma_{ed,ep}}{dx_B dQ^2}} \, . \; \; 
\end{align}
The advantage of considering the inclusive-normalized double ratio is that the nuclear PDF effects and their uncertainty is  canceled out so that $R_A$ is only sensitive to final-state modifications.
 We use the nNNPDF30nlo~\cite{Khalek:2022zqe} nuclear PDFs and NNFF10lo parametrization~\cite{Bertone:2017tyb} for the fragmentation functions (FFs)~\footnote{The uncertainty from  the fragmentation functions largely cancels in the ratio $R_A$. We have checked that FF uncertainties estimated using either NNFF1.0lo or NNFF1.0nlo remain subleading to the uncertainty of the cold nuclear matter parameters.}. To numerically solve Eqs.~(\ref{eq:main-eq-NS}), (\ref{eq:main-eq-S1}) and (\ref{eq:main-eq-S2}), 
 the initial condition is obtained by smearing the NLO calculation of the parton energy spectrum by a Gaussian wave packet. It is
 then evolved from $\mu_1=Q$ to $Q_0 =1$~GeV using the standard vacuum DGLAP equations.  
The in-medium RG evolves $\mu_2^2$ from $\chi(w) z\nu/L$ to $\xi^2$.
We take $\Lambda_{\rm QCD}=0.16$~GeV,  
the inverse range of the interaction squared  $\xi^2=0.12$~GeV$^2$ and the central value of the effective medium density parameter $\rho_G=0.4\; {\rm fm}^{-3}$. 
These values yield a quark transport coefficient $\hat{q}_F\approx 0.052$ GeV$^2$/fm at $\nu= 12$~GeV, consistent with existing SCET$_{\rm G}$ based mDGLAP applications~\cite{Li:2020rqj,Li:2020zbk} and compatible with values from  earlier SIDIS studies \cite{Arleo:2003jz,Wang:2009qb} within a factor of two.

\begin{figure}[ht!]
    \centering
\includegraphics[width=\columnwidth]{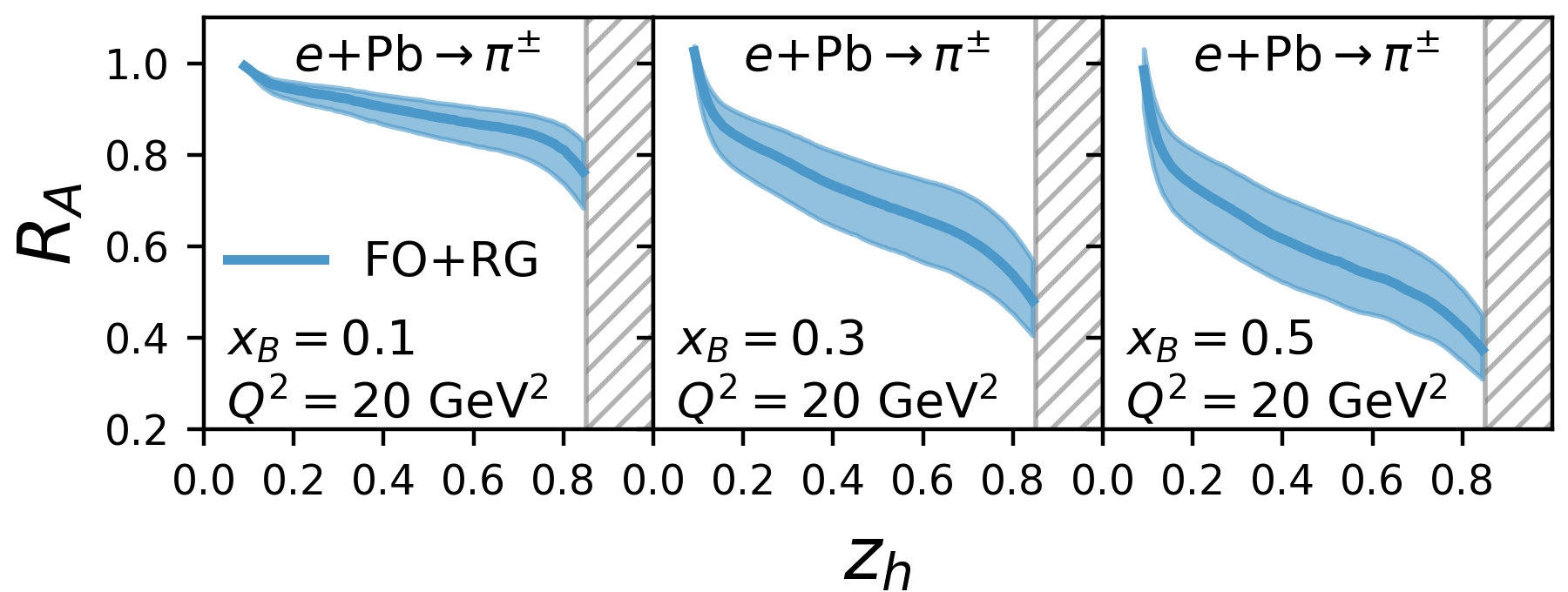}
\vspace*{-2em}
    \caption{Predictions for the modified pion fragmentation distribution at the EIC with Pb nucleus for three $(x_B, Q^2)$ combinations.}
    \label{fig:eic}
\end{figure}

The nuclear modification factor $R_A(z_h)$ calculated with the averaged HERMES kinematics $Q^2 = 2.25$ GeV$^2$ and $\nu=12$~GeV is compared to the HERMES data for $^{20}$Ne,  $^{84}$Kr and ${}^{131}$Xe targets \cite{HERMES:2007plz} in the top row of Fig.~\ref{fig:RA_HERMES_EMC}. Qualitatively, in-medium evolution shifts hadron spectra towards lower fragmentation fractions. Results including only RG contributions (red dashed lines) give a good description of $R_A$ from small to intermediate $z_h$, but lead to a suppression that is too strong at large $z_h$. The region close to $z_h = 1$ is dominated by 
threshold effects
where one should consider a different power counting and perform threshold resummation in both vacuum~\cite{Catani:1989ne,Anderle:2012rq} and in-medium calculations, so we exclude it from our comparison. In addition, hadrons produced with a small $z_h$ or $1-z_h$ and massive final states have short hadron formation times~\cite{Markert:2008jc}. If such time scales are smaller than the path length $L$, hadronic scattering, dissociation, and absorption in the nuclei have to be considered \cite{Accardi:2009qv}. For the reasons above, we will mainly discuss the intermediate $z_h$ region at HERMES.
Blue solid lines include the fixed order (FO) contribution from Eqs.~(\ref{eq:foG}), (\ref{eq:foNS}), (\ref{eq:foSS}) in the initial condition of the RG evolution, and the bands correspond to the density variation in the range $(\rho_G/1.5,  1.5 \rho_G)$. 
The FO correction improves the description of HERMES data at large $z_h$, but remains subleading to the RG evolution effect. 
The nuclear size dependence of $R_A$ is naturally explained with the same cold nuclear matter parameters. 

The cold nuclear matter parameters also give a good description of the EMC data \cite{EuropeanMuon:1991jmx}, which is the first measurement of nuclear modification to SIDIS hadron fragmentation. The measurements were performed for C, Cu, and Sn nuclei in a similar $x_B$ region, but with much higher $\langle Q^2\rangle\approx 11$ GeV$^2$ and, correspondingly, $\nu$, as compared to the HERMES experiment. We show in the bottom panel of Fig.~\ref{fig:RA_HERMES_EMC} the $R_A$ calculated at the averaged EMC kinematics using the same parameters that describe HERMES data. Even though the experimental error bars are larger for some of the measurements, there is a good agreement between the calculation and the data. Because the EMC data are taken at significantly larger $Q^2$ and $\nu$, the agreement demonstrates the predictive power and efficiency of this RG approach.

In Fig.~\ref{fig:eic}, we also project for modified pion fragmentation functions at the future electron-ion collider (EIC) for $e$Pb reactions at fixed $Q^2=20$ GeV$^2$ and various Bjorken $x_B$ values. We find that for $x_B>0.3$, where partons are less energetic in the nuclear rest frame, modifications become very large, consistent with existing predictions for heavy flavor and jets~\cite{Li:2020zbk,Li:2020rqj}. 
\section{Summary}
In the limit $Q/Q_0, E/\xi^2 L, \lambda_g\xi  \gg 1$ and to first order in the opacity of QCD matter, we performed a renormalization group analysis of medium effects for the SIDIS process on a nuclear target.
We derived a set of in-medium RG equations that resum the leading $\ln (E/\xi^2 L)$ terms from medium-induced emission and identified the corresponding fixed order corrections. We demonstrated that such logarithms are also treated in the phenomenological mDGLAP approach, which differs in the way of regulating the collinear divergences of in-medium bremsstrahlung and does not separate the fixed order contributions. Importantly, the new RG framework provides analytic insight into the salient features of parton showers responsible for the modification of hadron production in $eA$ which is not possible with numerical methods alone. It is a more efficient and systematically improvable way of treating the logarithmic enhancements in matter as compared to solving mDGLAP.

We applied the new method to study final-state cold nuclear matter (CNM) effects on single-hadron fragmentation and found that it gives a good description of the HERMES and EMC SIDIS data.  Predictions for the future EIC were also presented, where improved theoretical precision is especially important~\cite{AbdulKhalek:2021gbh}. The semi-analytic framework derived here can be generalized to initial-state CNM effects, such as the ones observed in Drell-Yan production in proton-nucleus collisions, and to heavy ion collisions. This work further benefits future QCD studies by guiding the incorporation of medium effects in Monte-Carlo event generators for the EIC, the Relativistic Heavy Ion Collider, and the Large Hadron Collider.

\section*{Acknowledgements}
The authors thank Duff Neill for the helpful discussion. This work is supported by the US DOE/SC under Contract No. 89233218CNA000001 and by the LDRD Program at LANL.

\bibliographystyle{elsarticle-num-names} 
\biboptions{square,numbers,compress}
\bibliography{ref}

\end{document}